# Momentum computed tomography of charged particles

Deyang Yu[*]

Institute of Modern Physics, Chinese Academy of Sciences, Lanzhou 730000, China.

[*]Email: d.yu@impcas.ac.cn

*The principle of the momentum computed tomography of charged particles is presented. It may be useful for momentum spectroscopy of various beam-matter interactions, especially when very intense beams are involved. It is able to collect the shower of charged particles with the $4\pi$ solid angle, and suitable for measuring the overall perspective of the arbitrary momentum distribution of the outgoing charged particles induced by arbitrary beams, especially when the other techniques are invalid. The extended collisional region, the strong field approximation and the case without magnetic field are discussed.*

The momentum distribution function of the outgoing ions or electrons carries the dynamical information in many kinds of reactions, such as in the particle or laser beam interacting with solid[1-4], cluster[5], molecule[6] or atom[7]. It is usually anisotropic in the momentum space because of the beam direction and sometimes the target orientation.

The traditional electrostatic and magnetic analyzers, like the spherical deflector analyzer[8-10], the 127° cylindrical deflector analyzer[11, 12], the cylindrical mirror analyzer[10, 13, 14], the plane mirror analyzer[15, 16], the retarding-field analyzer[17], the dipole magnetic spectrometer and the Thomson spectrometer[18], are always limited to small solid angles and at a certain aiming direction, which are not applicable to obtain the overall perspective of arbitrary momentum distribution.

In recent years, the velocity map imaging[19] and the so-called reaction microscope[20, 21] techniques have made significant progresses to measure the particle's momentum vectors in the $4\pi$ solid angle. Although the velocity map imaging technique has advantage in handling the ionic showers in the total-space of momentum, it is developed only for some special symmetric momentum distributions. On the other hand, the reaction microscope technique has unique advantage in single-collision which is able to obtain correlated momentum of two or three particles, but it has to measure the time-of-flight of the particles firstly and therefore limited to low count-rate experiments.

In this paper the principle of the computed tomography in the momentum space is presented, which is able to obtain the overall perspective of arbitrary momentum distributions of charged outgoing particles in the $4\pi$ solid angle, in spite of the arbitrary intensity and time structure of the reactions. It may be useful for momentum spectroscopy of various beam-matter interactions, especially when very intense beams are involved.

The scheme of the momentum computed tomography is shown in Figure 1. The collision point is located in a pair of paralleling electrostatic and magnetic fields. The charged particles produced in the collisions are driven by the fields and projected on the two-dimensional position-sensitive detectors, which also work as electrode plates. The uniform electrostatic and magnetic fields, as well as the positive ions are taken for instruction.

Let $F_{m,q}(v_x, v_y, v_z)$ represent the normalized velocity probability distribution function of the ions with mass $m$ and charge state $q$ (i.e., $\iiint_{-v_{max}}^{v_{max}} F_{m,q}(v_x, v_y, v_z) dv_x dv_y dv_z = 1$, while $v_{max}$ denotes the maximum possible speed of the ions), and let $\alpha_{m,q}$ denote its fraction (i.e., $\sum_{m,q} \alpha_{m,q} = 1$). Assuming the fields' direction is in the $z$-direction, and the detector is placed in the XOY plane with its center at the origin point, and the collision point (i.e., the starting point of the charged particles) is on the $z$-axle at the coordinate of $-d$. In a certain electrostatic field $E$ and a certain paralleling magnetic field $B$, consequently one will obtain a relevant normalized two-dimensional image distribution pattern $G_{E,B}(x, y)$ on the detector's surface (i.e., $\iint_{-r_{max}}^{r_{max}} G_{E,B}(x, y) dx dy = 1$, where $r_{max}$ denotes the maximum possible radius of the falling point). As physical density distributions, it is obvious that $F_{m,q}(v_x, v_y, v_z)$ and $G_{E,B}(x, y)$ are *finite*, *nonnegative*, *continuous*, *smooth* and *slow oscillating*. Formally, let the parametric operator $\hat{T}_{E,B}$ represent the effect of the fields, and $G_{E,B}(x, y)$ can be presented as follow:

$$G_{E,B}(x, y) = \hat{T}_{E,B} \sum_{m,q} \alpha_{m,q} F_{m,q}(v_x, v_y, v_z) \tag{1}$$

Although the momentum distribution depends only on the collision system (i.e., the beam and the target), each time of measurement produces a unique relating image density function $G_{E,B}(x, y)$ on the detector's surface with certain parameters $E$ and $B$, hence one will finally obtain a group of equations with different experimental parameters. As a result the unknown functions $F_{m,q}(v_x, v_y, v_z)$ and the relating coefficients $\alpha_{m,q}$ are able to be deduced from the equations group with sufficient times of measurement. Here, the explicit form of the equation (1) is shown as:

$$G_{E,B}(x, y) = \sum_{m,q} \alpha_{m,q} \int_{-v_{max}}^{v_{max}} J_{E,B} F_{m,q}(\tilde{v}_x, \tilde{v}_y, v_z) dv_z \tag{2a}$$

with

$$\tilde{v}_x = \frac{Bq}{2m} \left\{ x \cot\left[\frac{B}{2E}\left(\sqrt{v_z^2 + 2dqE/m} - v_z\right)\right] - y \right\} \tag{2b}$$

$$\tilde{v}_y = \frac{Bq}{2m} \left\{ x + y \cot\left[\frac{B}{2E}\left(\sqrt{v_z^2 + 2dqE/m} - v_z\right)\right] \right\} \tag{2c}$$

and

$$J_{E,B} = \left(\frac{Bq}{2m}\right)^2 \left\{ 1 + \cot^2\left[\frac{B}{2E}\left(\sqrt{v_z^2 + 2dqE/m} - v_z\right)\right] \right\} \tag{2d}$$

Let us consider an ion with mass $m$, charge state $q$, initial velocity $(v_x, v_y, v_z)$, and starting from the collision point $(0,0,-d)$. In the $z$-direction the ion is not affected by the magnetic field, and the time of flight of the particle is $t_f = m(\sqrt{v_z^2 + 2qdE/m} - v_z)/qE$. In the horizontal plane, governed by the magnetic field, the ion rotates with a constant period of $\tau = 2\pi m/qB$ with a constant cycling radius of $r = mv_\parallel/qB$, while $v_\parallel = \sqrt{v_x^2 + v_y^2}$. The projection of the rotating center on the detector's surface is at $(r\,v_y/v_\parallel, -r\,v_x/v_\parallel)$, and when it reaches the detector the rotating angle $\theta$ is $2\pi t_f/\tau$. Finally, the ion is projected on the detector's surface at the position $P$ which satisfies:

$$P: \begin{cases} x = \frac{m}{Bq}[v_x \sin\theta + v_y(1-\cos\theta)] \\ y = \frac{m}{Bq}[v_y \sin\theta - v_x(1-\cos\theta)] \end{cases} \qquad (3)$$

It implies that guided by the fields all of the ions represented by a curve $C$ in the momentum space will be projected on a same point $P$. And the curve satisfies:

$$C: \begin{cases} v_x = \frac{Bq}{2m}\left\{x\cot\left[\frac{B}{2E}\left(\sqrt{v_z^2 + 2dqE/m} - v_z\right)\right] - y\right\} \\ v_y = \frac{Bq}{2m}\left\{x + y\cot\left[\frac{B}{2E}\left(\sqrt{v_z^2 + 2dqE/m} - v_z\right)\right]\right\} \end{cases} \qquad (4)$$

Here the vertical velocity $v_z$ is the free parameter of the curve equation which varies from $-v_{max}$ to $v_{max}$. Therefore, every point on the detector's surface corresponds to a unique curve in the momentum space, and of course an element area on the detector surface is related to an element tube in the momentum space. Taking into account of the overlap of different charge states and masses, the integrated probability inside this element tube in the momentum space equals the probability inside the element area on the detector's surface. As a result, direct calculation lead to the equations (2).

From another point of view, the guiding fields will project all the ions with charge state $q$, mass m and velocity $(u_x, u_y, u_z)$ onto the detector's surface at the same point $(x_0, y_0)$, therefore the operator $\hat{T}_{E,B}$ of the fields satisfies:

$$\hat{T}_{E,B}\delta_{m,q}(v_x - u_x, v_y - u_y, v_z - u_z) = \delta(x - x_0, y - y_0) \qquad (5)$$

where $x_0$ and $y_0$ are defined by equation (3).

At the same time, any velocity distribution function can be represented as:

$$F_{m,q}(v_x, v_y, v_z) = \iiint_{-v_{max}}^{v_{max}} F_{m,q}(u_x, u_y, u_z)\delta_{m,q}(v_x - u_x, v_y - u_y, v_z - u_z)du_x du_y du_z \quad (6)$$

Therefore:

$$\hat{T}_{E,B}F_{m,q}(v_x, v_y, v_z) = \iiint_{-v_{max}}^{v_{max}} F_{m,q}(u_x, u_y, u_z)\delta(x - x_0, y - y_0)du_x du_y du_z \quad (7)$$

Taking into the account of the overlap of different charge states and the masses, again the calculation results in equations (2).

In many cases the collision region cannot be simply treated as a point. For examples, when a thin beam string passes through a diffusion gas target, or when a diffusion beam sputters a plane surface, the collision region is either a line segment or a planar area. In these cases, one can determinate the normalized luminosity distribution $L(x)$ or $L(x, y)$ by applying a sufficient strong electrostatic field (i.e., $E/d \to \infty$). Let $G'_{E,B}(x, y)$ denote the image pattern on the detector's surface which corresponds to the extended collision region, therefore it is the convolution of the point-like source image distribution pattern with the luminosity distribution due to the experimental layout (e.g., for the case of two-dimensional luminosity distribution):

$$G'_{E,B}(x, y) = (L * G_{E,B})(x, y) \quad (8)$$

For the instance of a two dimensional luminosity distribution, denoted by $L(\zeta, \eta)$, instead of equation (2b) and (2c), let:

$$\tilde{\tilde{v}}_x = \frac{Bq}{2m}\left\{(x - \zeta)\cot\left[\frac{B}{2E}\left(\sqrt{v_z^2 + 2dqE/m} - v_z\right)\right] - (y - \eta)\right\} \quad (9a)$$

$$\tilde{\tilde{v}}_y = \frac{Bq}{2m}\left\{(x - \zeta) + (y - \eta)\cot\left[\frac{B}{2E}\left(\sqrt{v_z^2 + 2dqE/m} - v_z\right)\right]\right\} \quad (9b)$$

Then the following formula holds:

$$G_{E,B}(x - \zeta, y - \eta) = \sum_{m,q} \alpha_{m,q} \int_{-v_{max}}^{v_{max}} F_{m,q}(\tilde{\tilde{v}}_x, \tilde{\tilde{v}}_y, v_z)J_{E_i,B_j}(v_z)dv_z \quad (9c)$$

Where $J_{E_i,B_j}(v_z)$ is still defined by the equation (2d). The equation (2a) is therefore instead by the following convolution form, which is the explicit form of the equation (8):

$$G'_{E,B}(x, y) = \int_{-\zeta_{max}}^{\zeta_{max}} \int_{-\eta_{max}}^{\eta_{max}} L(\zeta, \eta)G_{E,B}(x - \zeta, y - \eta)d\zeta d\eta \quad (9d)$$

Because the luminosity distribution is known in prior, the equation does not contain any more unknown information than equations (2), and then the unknown momentum distribution function can also be deduced with sufficient precision in spite of the extended collision region.

The present of the magnetic field has an advantage when both low energy ions and high energy electrons are measured at the same time (i.e., the two detectors are placed oppositely, for ions and electrons, respectively), such as in the ion-atom collisions[21]. In this case, the initial energy of the recoiled ions are typically less than 1eV while the initial energy of the outgoing electrons are typically several tens eV. Thus, the electrostatic field should be weak enough to avoid that all the ions are projected to the center region of the detector. Since the mass ratio between the ions and the electron is about $10^4$, an appropriate weak magnetic field will effectively restrain the electrons from going out of the detector.

Sometimes the magnetic field is unnecessary, e.g., when only low energy outgoing ions are measured. Consequently, the explicit form of the equation (1) is expressed as follow when $B = 0$:

$$G_E(x,y) = \sum_{m,q} \alpha_{m,q} \int_{-v_{max}}^{v_{max}} J_E F_{m,q}(\tilde{v}_x, \tilde{v}_y, v_z) dv_z \tag{10a}$$

with:

$$\tilde{v}_x = \frac{x}{2d}\left(\sqrt{v_z^2 + 2dqE/m} + v_z\right) \tag{10b}$$

$$\tilde{v}_y = \frac{y}{2d}\left(\sqrt{v_z^2 + 2dqE/m} + v_z\right) \tag{10c}$$

and

$$J_E = \left[\frac{1}{2d}\left(\sqrt{v_z^2 + 2dqE/m} + v_z\right)\right]^2 \tag{10d}$$

When the electrostatic field is sufficiently strong, as well as the flight distance is long enough, i.e., $v_z \ll \sqrt{2qdE/m}$, the initial velocity almost does not affect the time of flight. As a consequence, the vertical component of the velocity becomes dummy, and the momentum distribution is reduced to a two-dimensional function $F_{m,q}(v_x, v_y)$. In this case the integral transform disappears, and hence the transformation is greatly simplified:

$$G_{E,B}(x,y) = \sum_{m,q} \alpha_{m,q} J_{E,B} F_{m,q}(\tilde{v}_x, \tilde{v}_y) \tag{11a}$$

with

$$\tilde{v}_x = \frac{Bq}{2m}\left[x\cot(B\sqrt{dq/2mE}) - y\right] \tag{11b}$$

$$\tilde{v}_y = \frac{Bq}{2m}\left[x + y\cot(B\sqrt{dq/2mE})\right] \tag{11c}$$

and

$$J_{E,B} = \left(\frac{Bq}{2m}\right)^2 \left[1 + \cot^2(B\sqrt{dq/2mE})\right] \tag{11d}$$

The strong field approximation $v_z \ll \sqrt{2qdE/m}$ suggests that those charged particles have same horizontal velocity will hit on the same point at the detector. Owing to different velocity directions, the maximum discrepancy of the falling points is $mv_{max}^2/qE$, where $v_{max}$ is the maximum speed of the charged particles. Accordingly, the radius of the region of the falling points is about $v_{max}\sqrt{2md/qE}$. In the discretization and numerical solving procedure, if the detector is divided into $N \times N$ elemental areas, and then the maximum falling point discrepancy should be limited to a half spacing. The equivalent condition is $Eqd/N^2 > mv_{max}^2/2$, i.e., the ion's energy gain in the electrostatic field should be $N^2$ times bigger than its maximum original kinetic energy.

An even more simple case is the strong field approximation $v_z \ll \sqrt{2qdE/m}$ satisfies and simultaneously the magnetic field does not exist. In this case, the equations (10) and (11) are further reduced, respectively.

$$\sum_{m,q} \alpha_{m,q} J_E F_{m,q}(\tilde{v}_x, \tilde{v}_y) = G_E(x,y) \tag{12a}$$

with:

$$\tilde{v}_x = x\sqrt{qE/2md} \tag{12b}$$

$$\tilde{v}_y = y\sqrt{qE/2md} \tag{12c}$$

and

$$J_E = qE/2md \tag{12d}$$

When a continuous laser or an ion beam interacts with surfaces or another atomic, molecular or cluster beam, the measurement speed is limited by the maximal count rate of the detector, which typically reaches about $10^6$ Hz by the micro-channel plate detectors. For instance, in the case of an two-element substance target, $q_{max} = 3$, $N = 20$, and at least $10^4$ events (assuming averagely $10^5$ events) should be accumulated in each element area on the detector surface to produce smooth experimental image patterns, totally about $4.8 \times 10^9$ events (4800 seconds at the count rate of $10^6$ Hz) should be accumulated in all 120 times of measurement. In another case, when a very strong pulsed laser or ion beam (e.g., concerning to the high energy density researches) interacts with dense targets (e.g., solids, cluster beams), the CCD camera following micro-channel plate and phosphor screen should be utilized, which is widely used in the velocity map imaging technique[19]. In these cases the experimental duration depends on how many charged particles are produced by one pulse as well as the beam repeat frequency.

The engineering of the proposed momentum computed tomography can draw valuable lessons from the velocity map imaging[19] and the reaction microscope[20, 21] techniques, which have already been testified. Under the strong field approximation to obtain the two-dimensional momentum distributions is already feasible. However, in the three-dimension case, the relating high efficiency algorithms are desired.

The integral transform in equations (2) is a kind of the generalized Radon transform[22, 23]. To numerically solve the equations, the functions $F_{m,q}(v_x, v_y, v_z)$ and $G_{E,B}(x, y)$ should be discretized. If $N$ points are needed to represent it in any projection plane with sufficient precision, $N \times N \times N$ elemental cubes are needed in the momentum space, while each elemental cube is represented by its center coordinate and its average probability. For a substantial distribution function, $N = 10 \sim 20$ is usually enough to achieve a satisfied precision, while the non-grid points should be obtained by interpolation. Considering the charge states $q$ and the mass $m$, a group of additional unknown numbers $\alpha_{m,q}$ appear besides $N^3$ grid points for each charge state and each mass, and there are totally $n_m n_q (N^3 + 1)$ unknown numbers to be solved, where $n_m$ and $n_q$ represents the possible mass numbers and charge states, respectively. Accordingly, the detector's surface can be divided into $N \times N$ elemental areas, which are also represented by their center coordinates and average probabilities.

As mentioned above, for each discretized image point $(n, l)$, there exists a known corresponding curve $C_{(n,l)}$ in the momentum space. Furthermore, those elemental cubes which are passed through by the curve $C_{(n,l)}$ as well as the corresponding intercepts $\Delta v_z^{i,j,k}$ inside those elemental cubes are easily accessible. As a result, the equation (2) should be discretized as:

$$G_{E,B}(n, l) = \sum_{m,q} \alpha_{m,q} \sum_{i,j,k} F_{m,q}(i, j, k) J_{E,B}(k) \Delta v_z^{i,j,k} \tag{13}$$

The summation of $i$, $j$ and $k$ in the momentum space runs over those elemental cubes which are passed though by the corresponding curve $C_{(n,l)}$ of the discretized image point $(n, l)$. Now, the integral equations are reduced to algebraic equations. It should be noted that the equation (13) obtained by each time of measurement actually include $N^2$ independent equations due to the $N \times N$ subdivisions of the detector' surface.

Similarly, the discretized form of equation (8) of the finite collision region is:

$$G_{E,B}(n, l) = \sum_{m,q} \alpha_{m,q} \sum_{s,t} L(s, t) \sum_{i,j,k} F_{m,q}(i, j, k) J_{E,B}(k) \Delta v_z^{i,j,k} \tag{14}$$

where $s$ and $t$ denote the discretized coordinate of the luminosity, in which the same spacing are adopted according to the discretization of the image pattern on the detector's surface. Now the summation of $i$, $j$ and $k$ in the momentum space runs over those elemental cubes passed though by the corresponding curve of the discretized image point $(n - s, l - t)$, which is determined by equation (3).

When the strong field approximation satisfies, after discretization, for each mass and charge state there will be only $N^2$ unknown numbers in these two-dimensional cases, instead of $N^3$ in three-dimensional case. Because each time of the measurement set up $N^2$ independent equations, only a few measurements are need to determine the distribution functions $F_{m,q}(v_x, v_y)$ and the corresponding factors $\alpha_{m,q}$.

**Acknowledgements**

The Author thanks Yingli Xue, Pingxiao Wang, Jie Yang, Xiaohong Cai, Xurong Chen, Yipan Guo and Chonghong Zhang for carefully reading the manuscript. Special thanks to Dieter. H. H. Hoffmann and Zhongkai. Li who checked the physical and mathematical derivation. This work was financially supported by the National Natural

Science Foundation of China under Grant No. 11275240 and the National Basic Research Program of China under Grant No. 2010CB832901.

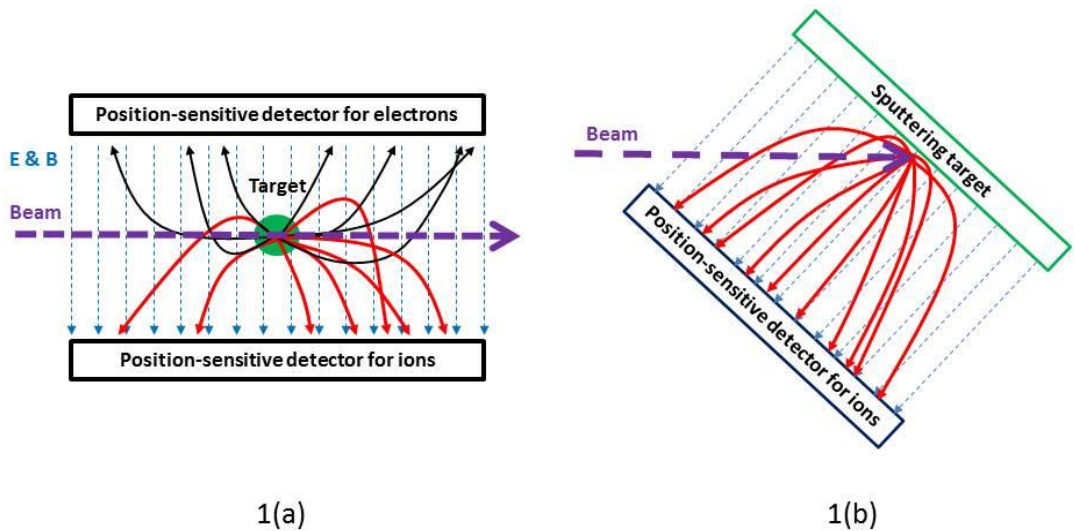

**Figure 1** The diagrammatic scheme of the momentum computed tomography. In figure (1a) a jet target perpendicularly passes through the paper and is collided by a beam, and in figure (1b) a solid target is embedded into an electrode plate and sputtered by a beam. The outgoing ions (as well as the electrons) with their initial momentum are driven by the electrostatic and the magnetic fields, and projected on the two-dimensional position sensitive detectors. Then a distribution pattern owing to certain fields' parameters will be produced. By changing the fields' parameters, the fraction of charge states and the relating velocity distribution functions can be deduced.